\documentclass[aps,pre,twocolumn,10pt,showpacs,superscriptaddress,preprintnumbers,amsmath,amssymb]{revtex4-1} 
\usepackage{graphicx,dcolumn,bm,times,mathtools}

\begin{document}

\title{Memory function of turbulent fluctuations in soft-mode turbulence}

\author{Takayuki Narumi}
\email{narumi@athena.ap.kyushu-u.ac.jp}
\affiliation{Department of Applied Quantum Physics and Nuclear Engineering, Kyushu University, Fukuoka 819-0395, Japan}
\author{Junichi Yoshitani}
\affiliation{Department of Applied Quantum Physics and Nuclear Engineering, Kyushu University, Fukuoka 819-0395, Japan}
\author{Masaru Suzuki}
\affiliation{Department of Applied Quantum Physics and Nuclear Engineering, Kyushu University, Fukuoka 819-0395, Japan}
\author{Yoshiki Hidaka}
\email{hidaka@ap.kyushu-u.ac.jp}
\affiliation{Department of Applied Quantum Physics and Nuclear Engineering, Kyushu University, Fukuoka 819-0395, Japan}
\author{\\Fahrudin Nugroho}
\affiliation{Department of Applied Quantum Physics and Nuclear Engineering, Kyushu University, Fukuoka 819-0395, Japan}
\affiliation{Physics Department, Gadjah Mada University, Yogyakarta 55281, Indonesia}
\author{Tomoyuki Nagaya}
\affiliation{Department of Electrical and Electronic Engineering, Oita University, Oita 870-1192, Japan}
\author{Shoichi Kai}
\affiliation{Department of Applied Quantum Physics and Nuclear Engineering, Kyushu University, Fukuoka 819-0395, Japan}

\date{\today}

\begin{abstract} %
Modal relaxation dynamics has been observed experimentally to clarify statistical-physical properties of soft-mode turbulence,
the spatiotemporal chaos observed in homeotropically aligned nematic liquid crystals.
We found a dual structure, dynamical crossover associated with violation of time-reversal invariance,
the corresponding time scales satisfying a dynamical scaling law.
To specify the origin of the dual structure,
the memory function due to non-thermal fluctuations has been defined by a projection-operator method
and obtained numerically using experimental results.
The results of the memory function suggest that the non-thermal fluctuations can be divided into Markov and non-Markov contributions,
the latter is called the turbulent fluctuation (TF).
Consequently, the relaxation dynamics is separated into three characteristic stages:
bare-friction, early, and late stages.
If the dissipation due to TFs dominates over that of the Markov contribution,
the bare-friction stage contracts;
the early and late stages then configure the dual structure.
The memory effect due to TFs results in the time-reversible relaxation at the early stage, 
and the disappearance of the memory by turbulent mixing 
leads to a simple exponential relaxation at the late stage.
Furthermore, the memory effect due to TFs is shown 
to originate from characteristic spatial coherency called the patch structure.
\end{abstract}


\maketitle

\section{Introduction}
Weak nonlinearity can generate spatial and temporal disorders
in systems where the number of effective degrees of freedom increases with increasing system size.
A phenomenon triggered by weak nonlinearity in such high-dimensional systems is called spatiotemporal chaos,
in contrast with chaos where unpredictable behavior emerges 
from few degrees of freedom in a deterministic way.
Theoretical work on the spatiotemporal chaos
 (Refs.~\cite{Kai1991_Proceeding, Cross1993} and references therein)
has covered topics such as the complex Ginzburg--Landau equation, 
the Kuramoto--Sivashinsky turbulence (KST), the Nikolaevskii turbulence, 
and coupled map lattices.
An outstanding feature of chaos and turbulence is its dual structure,
where the dynamics is separated into an \textit{initial} regime corresponding to deterministic short orbits 
and a \textit{final} regime corresponding to stochastic long orbits.
Mori and Okamura, for example, have theoretically studied 1D-KST
and revealed the dual structure in turbulent mixing by numerical analysis \cite{Mori2009}.

Convection systems have been experimentally investigated to study nonlinearity.
An example is electrohydrodynamic convection observed in the nematic liquid crystal
and controlled by an ac voltage applied to system.
One can easily achieve accessible characteristic length and time scales 
in experiments of electroconvection compared with the Rayleigh--B\'{e}nard convection.
Another advantage of investigating nematics is the ease 
in controlling the anisotropy to study symmetry properties.
There are two types of layer alignment in nematic liquid crystals;
one is planar alignment in which the director aligns parallel to substrates ($x$-direction),
and the other is homeotropic alignment in which the director aligns perpendicular to substrates ($z$-direction).
Rubbing along the $x$-direction of a substrate's surface ($x$--$y$ plane) 
produces planar systems,
and intrinsically breaks the rotational symmetry.
In homeotropic systems, by contrast, 
the rotational symmetry in the $x$--$y$ plane remains.
With a sufficiently strong applied ac voltage, magnitude $V$ and fixed frequency $f$,
the Fr\'{e}edericksz transition, occurring at a certain threshold voltage $V_{\text{F}}$,
spontaneously breaks the rotational symmetry \cite{Freedericksz1933};
the transition is accompanied by the excitation of the Nambu--Goldstone modes \cite{Hertrich1992, Richter1995_EL, Tribelsky1996}.
By further increasing $V$, 
electrohydrodynamic convection occurs at $V_{\text{c}}$.
The nonlinear coupling between the convective and Nambu--Goldstone modes generates 
a pattern that is both spatially and temporally disordered.
The experimentally observed phenomenon, called soft-mode turbulence (SMT),
is an example of spatiotemporal chaos \cite{Kai1996, Hidaka1997_PRE, Hidaka1997_JPSJ}.

In our preceding study, we have observed SMT relaxation by measuring a temporal autocorrelation function
and reported that the relaxation is well fitted by a compressed exponential function \cite{Nugroho_SMTtotal_2012}.
Because the compressed exponential is employed to describe the dynamics of jammed systems \cite{Bouchaud2008},
we remarked on the similarity between SMT and glass forming liquids.
The nonlinearity in the dynamics of the latter originates from dynamic coherency in some regions.
In fact, spatiotemporal fluctuating cooperative regions have been observed 
as dynamical heterogeneities near the glass transition point \cite{Sillescu1999, Ediger2000, Richert2002, Berthier2011_book}.
A characteristic length for these cooperative regions increases as the glass transition point is approached.
In SMT, on the other hand, patch domains exist in which convective rolls align in a unique orientation \cite{Tamura2002, Hidaka2006, Tamura2006}.
The characteristic size $\xi$ of a patch domain is several times longer than the typical size of convective rolls
and decreases with the distance from SMT onset; $\xi\sim\varepsilon^{-1/2}$ with control parameter $\varepsilon$.
We have therefore concluded that 
the SMT patch domains behave like the cooperative regions in glass forming liquids
and the coherent motion in the domains generates non-exponential relaxation.
To study SMT dynamics in detail,
the temporal correlations of each wave number, \textit{modal} autocorrelation functions, are suitable.
Our previous study focused on the \textit{net} autocorrelation function consisting of the entire wave-number information.
Here, we investigate the modal relaxation dynamics to specify statistical physics of SMT.

\section{Experiment}  \label{sec:experiment}
We study a 2D pattern dynamics of SMT observed in a homeotropic alignment of nematic liquid crystals.  
This study follows a standard setup \cite{Kai1996, Tamura2001, Anugraha2008}.
The space between two parallel glass plates, spaced $27~\mu$m apart,
was filled with the nematic liquid crystal,
$N$--(4--Methoxybenzilidene)--4--buthylaniline (MBBA).
The plate surfaces were coated with transparent electrodes,
made of indium tin oxide (ITO) with a circular cross-section of radius $13$ mm.
To obtain homeotropic alignment,
the surfaces was covered by a surfactant, 
$N$, $N$--dimethyl--$N$--octadecyl--3--aminopropyl-trimethoxysilyl chloride 50\% (DMOAP). 
The values of the dielectric constant and electric conductivity parallel to the director were
$\epsilon_{\parallel}=6.25$ and $\sigma_{\parallel}=1.17 \times 10^{-7}~\Omega^{-1}\text{m}^{-1}$, respectively.
Denoting the dielectric constant perpendicular to the director by $\epsilon_{\perp}$,
the dielectric constant anisotropy $\epsilon_{a}=\epsilon_{\parallel}-\epsilon_{\perp}$ is found to be negative.
An ac voltage $V(t)=\sqrt{2}V\cos(2\pi f t)$ was applied to the sample.
For a control parameter,
we employ a normalized voltage $\varepsilon=\left(V/V_{c}\right)^{2}-1$,
where $V_{c}$ denotes the threshold voltage for electroconvection having value $7.78 \pm 0.05$~V. 
We show results for $\varepsilon = 0.025, 0.050, 0.075, 0.10, 0.20, 0.30,$ and $0.40$,
where $\varepsilon$ has a margin of error of  $\pm~0.013$.
Another control parameter was the frequency $f$ of the ac voltage.
Two patterns of SMT arise; oblique roll in $f < f_{\text{L}}$ and normal roll in $f > f_{\text{L}}$,
where $f_{\text{L}}$ denotes the Lifshitz frequency \cite{Kai1996,Hidaka1997_PRE}.
We set the frequency $f=100$~Hz well below $f_{\text{L}}$.
The temperature was regulated to $30.00\pm 0.05~^{\circ}\text{C}$.
Before each sampling,
we waited for $10$~min at fixed voltage $V_{\text{w}}$ and then a further $10$~min at the set $V$ to get a desired $\varepsilon$ value, 
where $V_{\text{F}}<V_{\text{w}}<V_{\text{c}}$. 
The waiting time is sufficiently long for systems to achieve steady state.

\begin{figure}[!t]
\begin{center}
\includegraphics[width=0.99\linewidth]{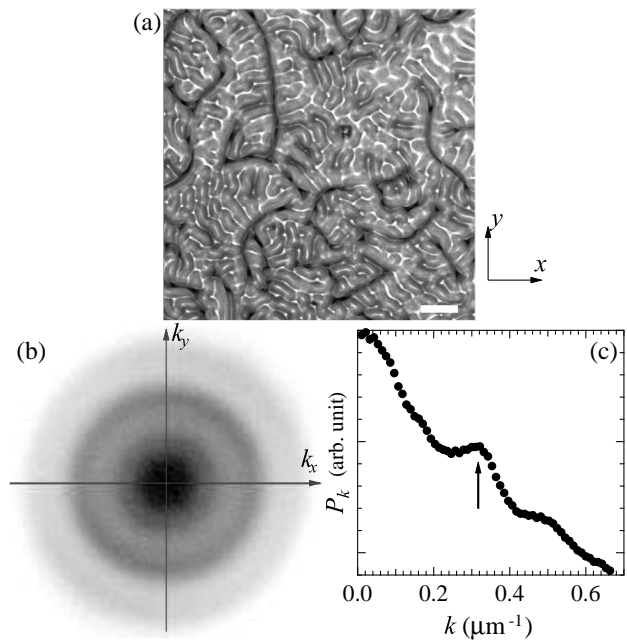}
\end{center}
	\caption{
	Static information of our experiment at $\varepsilon=0.1$.
	(a) A typical snapshot of SMT with the white scale bar indicating $100$~$\mu$m.
		Each bright line indicates upward flow.
	(b) Magnitude tints of the spatial power spectrum $P_{\bm{k}}$ in the Fourier space,
		where shading corresponds to the $P_{\bm{k}}$ value.
	(c) Plot of the power spectrum $P_{k}$ as a function of the radial wave number $k = |\bm{k}|$.
		The peak, marked by an arrow, corresponds to the fundamental period in a convective rolling.
	}
\label{fig:experiment}
\end{figure}

The electroconvection pattern was observed by a microscope (ECLIPSE E600POL, Nikon Corporation, Tokyo)
and was captured by a high-speed camera (HAS220, DITECT Co.~Ltd., Tokyo) that can successively take $4200$ frames.
The numbers of frames per second were $5$ for $\varepsilon<0.1$ and $10$ otherwise.
A typical two-dimensional image is shown in Fig.~\ref{fig:experiment} (a).
The measurement area was $830 \times 830$ $\mu\text{m}^{2}$ ($450 \times 450$ pixels). 
The transmitted light intensity $I(\bm{x},t)$ at each pixel was digitized into $8$-bit (i.e., $256$-level) information,
where a series of pattern analyzing was processed according to Ref.~\cite{Nagaya1999}. 

With angle brackets signifying the long-time average in the steady state,
the temporal correlation of two functions is defined as
\begin{equation}
	\left<f(t+\tau)g(t)\right> = \lim_{T\to \infty}\frac{1}{2T}\int_{-T}^{T}{\rm d}t f(t+\tau)g(t).
	\label{def_LTaverage}
\end{equation}
We employ the Fourier transform of the fluctuation $\Delta I(\bm{x},t) = I(\bm{x},t) - \left<I(\bm{x},t)\right>$ 
as the gross variable $u_{\bm{k}}(t)$;
\begin{equation} 
	u_{\bm{k}}(t) \coloneqq \int {\rm d} \bm{x} \Delta I(\bm{x},t)e^{\text{i} \bm{k} \cdot \bm{x}}, 
\end{equation}
where $\text{i}=\sqrt{-1}$ and the integral range is over the entire system domain.
In isotropic systems, 
it is sufficient to study $u_{k}(t)$ \cite{Nagaya2000},
where $k$ denotes the radial wave number; $k=|\bm{k}|$.
We focus on the normalized modal time-correlation function of $k$;
\begin{equation}
	\hat{U}_{k}(\tau) : = \left<u_{k}(t+\tau)u_{k}^{*}(t)\right>P_{k}^{-1},
	\label{def_corr_func}
\end{equation}
where the asterisk denotes the complex conjugate operation
and $P_{k} = \left<|u_{k}(t)|^{2}\right>$ denotes the spatial power spectrum.
Note that $\hat{U}_{k}(\tau)$ is a real number due to translational symmetry and isotropy. 
The spatial power spectra as functions of the wave number $\bm{k}$ and $k=|\bm{k}|$ 
are illustrated in Fig.~\ref{fig:experiment} (b) and (c), respectively.
SMT isotropy is clearly reflected in Fig.~\ref{fig:experiment} (b).
Although experimental parameters in Ref.~\cite{Nagaya2000} are slightly different from ours, 
the $P_{k}$ profiles agree.
A clear peak exists at $k_{\text{peak}}\simeq 0.321~\mu\text{m}^{-1}$ for $\varepsilon=0.10$
and its corresponding length $\lambda_{\text{peak}} = 2\pi / k_{\text{peak}}$ is $19.6~\mu\text{m}$ that is a half of $\lambda_{0}$, 
where $\lambda_{0}=\lambda_{0}(\varepsilon)$ denotes the length of a pair of electroconvections.
The peak wavenumbers $k_{\text{peak}}$ for other $\varepsilon$ values are similar.
The wave number $k$ is normalized by $\lambda_{0}$ as $\hat{k}=k \lambda_{0}/2\pi$.

\section{Result and Discussion} \label{sec:RandD}

\subsection{Dual relaxation}

\begin{figure}[!b]
\begin{center}
\includegraphics[width=0.9\linewidth]{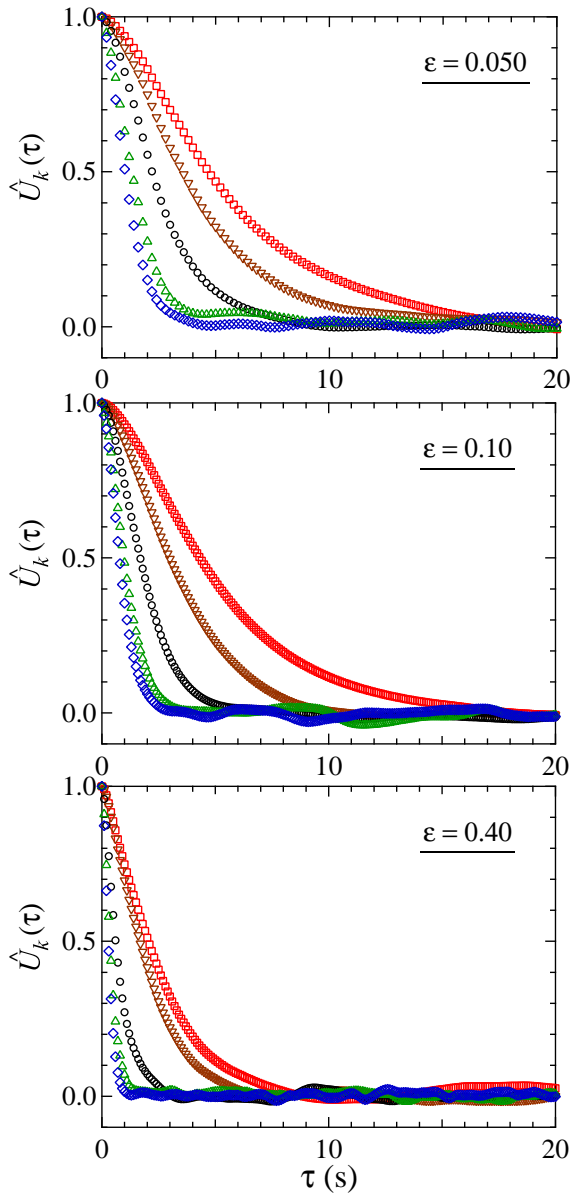}
\end{center}
	\caption{(color online)
		Plot of the modal correlation functions at $\varepsilon=0.050$ (top), $0.10$ (middle), and $0.40$ (bottom).
		The wave numbers $\hat{k}$ are $0.76$ (red square), $1.0$ (brown inverted triangle), 
		$2.0$ (black circle), $3.0$ (green triangle), and $4.0$ (blue diamond).
	}
\label{fig:cor}
\end{figure}

\begin{figure}[!b]
\begin{center}
\includegraphics[width=0.9\linewidth]{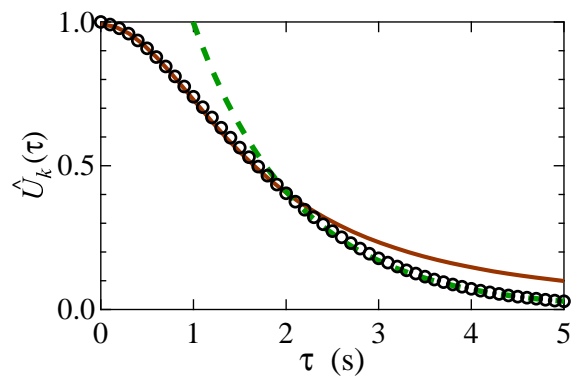}
\end{center}
	\caption{(color online)
		The normalized modal autocorrelation function of $\varepsilon=0.1$ and $\hat{k}=2.0$,
		which is the same as shown in Fig.~\ref{fig:cor}.
		The algebraic decay (\ref{algebraic_decay}) (brown solid line) well describes the dynamics in the early stage,
		but a simple exponential (\ref{exponential_decay}) (green dashed line) is better at the late stage.
	}
\label{fig:dualstructure}
\end{figure}

The modal autocorrelation function $\hat{U}_{k}(t)$ obtained experimentally is shown in Fig.~\ref{fig:cor}.
In the short-time regime, 
the relaxation is not described by a simple exponential function.
Instead, we found dual relaxation as evident in Fig.~\ref{fig:dualstructure}.
In the early stage \cite{Note1},
the correlation function follows an algebraic decay form;
\begin{equation}
	\hat{U}_{k}(\tau) \propto 1- \left(\tau/ \tau^{(\text{a})}_{k} \right)^{2}.
	\label{algebraic_decay}
\end{equation}
In contrast,
the relaxation in the late stage is well described by exponential decay;
\begin{equation}
	\hat{U}_{k}(\tau) \propto \exp\left[-\tau\left/ \tau^{(\text{e})}_{k}\right.\right].
	\label{exponential_decay}
\end{equation}
\begin{figure}[!t]
\begin{center}
\includegraphics[width=0.91\linewidth]{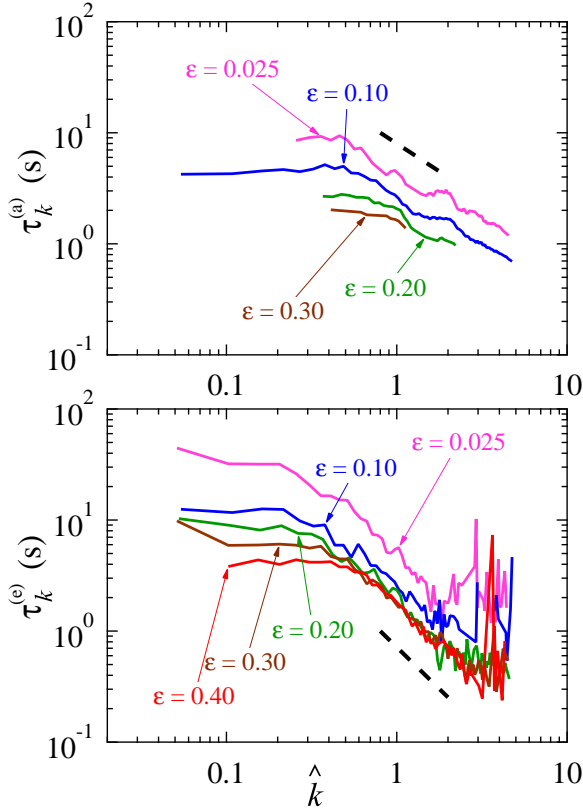}
\end{center}
	\caption{(color online)
		Log-log plots of time scales $\tau^{(\text{a})}_{k}$ (top) and $\tau^{(\text{e})}_{k}$ (bottom)
		for several control parameters;
		$\varepsilon=0.025$ (pink), $0.10$ (blue), $0.20$ (green), $0.30$ (brown), and $0.40$ (red).
		The dashed lines plot the power law with exponents 
		$z_{\text{a}}=1.0$ (top panel) and $z_{\text{e}}=1.5$ (bottom panel).
	}
\label{fig:tauAandE}
\end{figure}

Figure \ref{fig:tauAandE} presents the characteristic time scales at each stage for the dual structure. 
The time scales are almost constant at small wave numbers.
Note that the results at small $\hat{k}$ may be affected by limitations in sample averaging.
At large wave numbers, 
the time scales of the dual relaxation satisfy dynamic scaling laws;
\begin{equation}
	\tau^{(\text{a})}_{k} \propto \hat{k}^{-z_{\text{a}}}~~,~~~~ \tau^{(\text{e})}_{k} \propto \hat{k}^{-z_{\text{e}}}  
	\label{D_exponent}
\end{equation}
with dynamical exponents $z_{\text{a}}\simeq 1.0$ and $z_{\text{e}}\simeq 1.5$ regardless of $\varepsilon$.
Although Mori and Okamura have reported that the dynamical scaling exponents obtained theoretically 
in 1D-KST are $z_{\text{a}}=1$ and $z_{\text{e}}=2$ \cite{Mori2009},
we cannot directly compare our results with their theoretical work 
because the dynamical exponent depends on spatial dimensionality.
Indeed, dynamic scaling for several turbulence models depends on the dimensionality \cite{Forster1977, Yakhot1981, Kardar1986}.

To elucidate the dual structure of SMT theoretically,
we have taken account of an experimental fact
that the spatiotemporal disorder accompanied by the patch structure can be represented 
as non-thermal fluctuations \cite{Tamura2002, Hidaka2010}.
The projection-operator method for turbulence  \cite{Mori2001} is one of the most suitable methods
to manage such non-thermal fluctuations;
its mathematical procedure derives an evolution equation containing the memory effect 
that is equivalent to the autocorrelation of fluctuations (see Appendix \ref{appendix:NPOF}).
Thus, in the rest of the paper,
analysis for the memory functions plays a central role to study a relationship between the dual structure and the non-thermal fluctuations.

\subsection{Memory effect due to non-thermal fluctuations} \label{sec:mem}
We assume the modal elements $\{u_{\bm{k}}\}$ form a complete set of macroscopic variables in SMT \cite{Note2}.
According to the projection-operator formalism,
the evolution equation in SMT is represented as
\begin{equation}
	\frac{\partial \hat{U}_{k}(\tau)}{\partial \tau} = -\int_{0}^{\tau}{\rm d}\tau^{\prime}\Gamma_{k}^{\prime}(\tau-\tau^{\prime})\hat{U}_{k}(\tau^{\prime}),
	\label{EOM_modalcorr_prime}
\end{equation}
where $\Gamma^{\prime}_{k}(\tau)$ denotes the memory function that results from the non-thermal fluctuations.
Here, translational symmetry reduces the temporal correlation of the modal elements to
\begin{equation}
	\left<u_{\bm{k}}(t+\tau)u_{\bm{k}^{\prime}}^{*}(t)\right> = \delta_{\bm{k},\bm{k}^{\prime}}\left<u_{\bm{k}}(t+\tau)u_{\bm{k}}^{*}(t)\right>
\end{equation}
and the mechanical coefficients $\omega_{\bm{k}\bm{k}}$ are zero by definition.

\begin{figure}[!t]
\begin{center}
\includegraphics[width=0.99\linewidth]{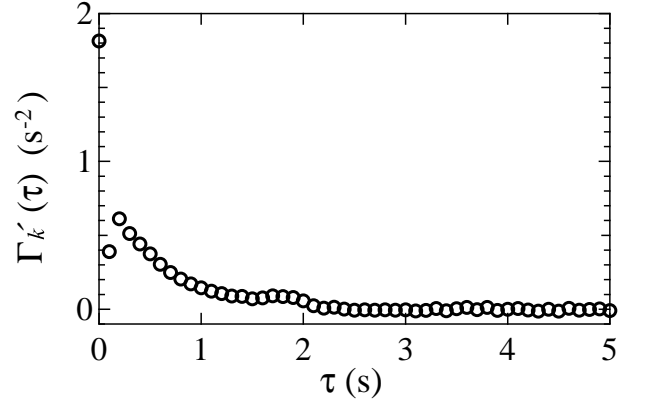}
\end{center}
	\caption{
		The memory function $\Gamma^{\prime}_{k}(\tau)$ characterized by Eq.~(\ref{EOM_modalcorr_prime}).
		This is the result of $\varepsilon=0.1$ and $\hat{k}=2.0$ as an example.
	}
\label{fig:memboth_0100}
\end{figure}

Using experimental results for the modal correlation function,
one can numerically solve Eq.~(\ref{EOM_modalcorr_prime}) to obtain the memory function.
The memory function $\Gamma^{\prime}_{k}(\tau)$ has a sharp peak at $\tau=0$,
as evident in Fig.~\ref{fig:memboth_0100},
implying that the non-thermal fluctuations can be separated
into Markov and non-Markov contributions.
Therefore, with
\begin{equation}
	\Gamma_{k}^{\prime}(\tau) = 2\gamma^{(0)}_{k}\delta(\tau)+\Gamma_{k}(\tau),
	\label{memory_function_SMT}
\end{equation}
where $\gamma^{(0)}_{k}$ denotes the bare friction due to the Markov contribution of the non-thermal fluctuations
and $\Gamma_{k}(\tau)$ the memory function due to the non-Markov contribution,
the evolution equation (\ref{EOM_modalcorr_prime}) reduces to
\begin{equation}
	\frac{\partial \hat{U}_{k}(\tau)}{\partial \tau} =  -\gamma^{(0)}_{k}\hat{U}_{k}(\tau)-\int_{0}^{\tau}{\rm d}\tau^{\prime}\Gamma_{k}(\tau-\tau^{\prime})\hat{U}_{k}(\tau^{\prime}).
	\label{EOM_modalcorr}
\end{equation}
To emphasize the transport due to the turbulent-like dynamics,
the non-Markov contribution to the non-thermal fluctuations is here called the \textit{turbulent fluctuations} (TFs) in SMT.
One can define a characteristic time scale for the memory function due to TFs as
\begin{equation}
	\tau^{(\Gamma)}_{k} : = \frac{1}{\Gamma_{k}(0)}\int_{0}^{\infty}{\rm d}\tau \Gamma_{k}(\tau),
	\label{tau_Gamma}
\end{equation}
within which the memory effects due to TFs are alive.

\begin{figure}[!t]
\begin{center}
\includegraphics[width=0.99\linewidth]{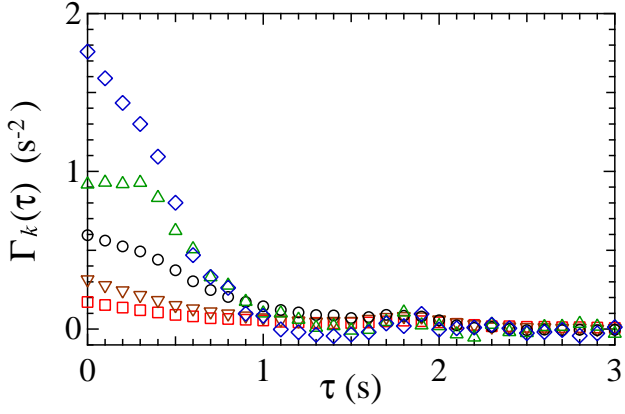}
\end{center}
	\caption{(color online)
		Time-dependence of the memory function for $\varepsilon=0.10$ at several fixed wave numbers:
		$\hat{k}=0.76$ (red square), $1.0$ (brown inverted triangle), $2.0$ (black circle), $3.0$ (green triangle), and $4.0$ (blue diamond).
	}
\label{fig:memNL}
\end{figure}

As seen in Fig.~\ref{fig:memNL}, the memory function $\Gamma_{k}(t)$ caused by TFs 
has non-negligible magnitudes and time scales over a range of wave numbers in the early stage;
therefore, SMT dynamics is regarded as being non-Markovian due to TFs.
The memory has a small peak around $\tau\simeq 2$ s.
We believe that this is due to a residual of the mechanical coefficient appearing in the projection-operator method,
despite the coefficient being theoretically zero.

Each equation of the dual relaxation is elucidated by the evolution equation (\ref{EOM_modalcorr}).
First, let us consider the dynamics at the late stage, $\tau^{(\Gamma)}_{k} \ll \tau$.
The frequency-dependent friction $\Gamma_{k,\omega}$ caused by TFs is defined 
as the Fourier-Laplace transformation of the memory function;
\begin{equation}
	\Gamma_{k,\omega} \coloneqq \int_{0}^{\infty}{\rm d}\tau\Gamma_{k}(\tau)e^{\text{i}\omega \tau}.
\end{equation}
At the late stage,
friction can be regarded as static, 
\begin{equation}
	\gamma^{(\Gamma)}_{k} \coloneqq \Gamma_{k,\omega=0};
\end{equation}
hence, the memory effect is approximated by a delta function
\begin{equation}
	\Gamma_{k}(\tau) \simeq 2\gamma^{(\Gamma)}_{k}\delta(\tau).
	\label{approximation_latestage}
\end{equation}
The modal relaxation dynamics is thus obtained as a simple exponential function (\ref{exponential_decay})
with the characteristic time 
\begin{equation}
	\tau^{(\text{e})}_{k} = \left(\gamma^{(0)}_{k} + \gamma^{(\Gamma)}_{k}\right)^{-1}.
	\label{theory_tau_e}
\end{equation}
Next, at $\tau \ll \tau^{(\Gamma)}_{k}$, an approximate solution of Eq.~(\ref{EOM_modalcorr}) is
\begin{equation}
	\hat{U}_{k}(\tau) = 1 - \frac{\gamma^{(\Gamma)}_{k}}{\tau^{(\Gamma)}_{k}}\int_{0}^{\tau} {\rm d}\tau^{\prime} \left( \tilde{\gamma}_{k}\tau^{(\Gamma)}_{k} + \tau^{\prime} \right) +\mathcal{O}(\tau^{3})
\end{equation}
with $\tilde{\gamma}_{k}=\gamma^{(0)}_{k}/\gamma^{(\Gamma)}_{k}$.
Thus, if $\tilde{\gamma}_{k}\ll 1$, 
the time range can be separated into $\tau \ll \tilde{\gamma}_{k}\tau^{(\Gamma)}_{k}$ and $\tilde{\gamma}_{k}\tau^{(\Gamma)}_{k} \ll \tau \ll \tau^{(\Gamma)}_{k}$.
In the shorter time regime $\tau \ll \tilde{\gamma}_{k}\tau^{(\Gamma)}_{k}$, 
called the bare-friction stage, modal relaxation decays linearly with slope $\gamma^{(0)}_{k}$.
In contrast, it reduces to time-reversible algebraic decay
in the longer time regime $\tilde{\gamma}_{k}\tau^{(\Gamma)}_{k} \ll \tau \ll \tau^{(\Gamma)}_{k}$,
i.e., the early stage of the dual structure with characteristic time 
\begin{equation}
	\tau^{(\text{a})}_{k} =\sqrt{2 \tau^{(\Gamma)}_{k}\left/ \gamma^{(\Gamma)}_{k} \right. }.
	\label{theory_tau_a}
\end{equation}
Therefore, a small ratio $\tilde{\gamma}_{k}$ of the friction coefficients 
is a necessary requirement for the appearance of dual relaxation in SMT.
The characteristic stages are summarized in Table \ref{table:stage}.

\begin{table*}
\caption{Characteristic stages of modal relaxation dynamics in soft-mode turbulence.}
\begin{tabular}{lrrll}
\hline
\hline
\multicolumn{1}{c}{stage} & \multicolumn{3}{c}{time range} &  \multicolumn{1}{c}{decay form}  \\
\hline
~~Bare-friction stage~~ & & $ \tau$ & $ \ll \tilde{\gamma}_{k}\tau^{(\Gamma)}_{k}$ &   ~~Linear decay caused by the Markov fluctuations~~\\
~~Early stage~~ & $\tilde{\gamma}_{k}\tau^{(\Gamma)}_{k}$ & $ \ll \tau$ & $ \ll \tau^{(\Gamma)}_{k}$& ~~Time-reversible decay that originates from memory due to TFs [Eq.~(\ref{algebraic_decay})]~~\\
~~Late stage~~  & $\tau^{(\Gamma)}_{k}$ & $\ll \tau$ & & ~~Simple exponential decay after turbulent mixing [Eq.~(\ref{exponential_decay})]~~\\
\hline
\hline
\end{tabular}
\label{table:stage}
\end{table*}
%

\subsection{Characteristic time scales}
A characteristic time scale for the modal autocorrelation function can be defined by
\begin{equation} 
	\tau^{(\text{U})}_{k} : = \int_{0}^{\infty} {\rm d}\tau \hat{U}_{k}(\tau). 
	\label{tau_U}
\end{equation}
This is the relaxation time due to non-thermal fluctuations.
In contrast, the time scale $1/\gamma^{(0)}_{k}$ denotes the relaxation time due to Markov fluctuations only.
From the evolution equation \eqref{EOM_modalcorr},
$\tau^{(\text{U})}_{k}$ is exactly represented by the friction coefficients,
\begin{equation}
	\tau^{(\text{U})}_{k} 	= \left(\gamma^{(0)}_{k} + \gamma^{(\Gamma)}_{k}\right)^{-1}.
	\label{relation_tau}
\end{equation}
Therefore, $\tau^{(\text{U})}_{k}$ is nearly equal to $\tau^{(\text{e})}_{k}$ according to Eq.~\eqref{theory_tau_e};
its accuracy has been confirmed within the margins of numerical error,
indicating that the approximation \eqref{approximation_latestage} is reasonable.

\begin{figure}[!t]
\begin{center}
\includegraphics[width=0.99\linewidth]{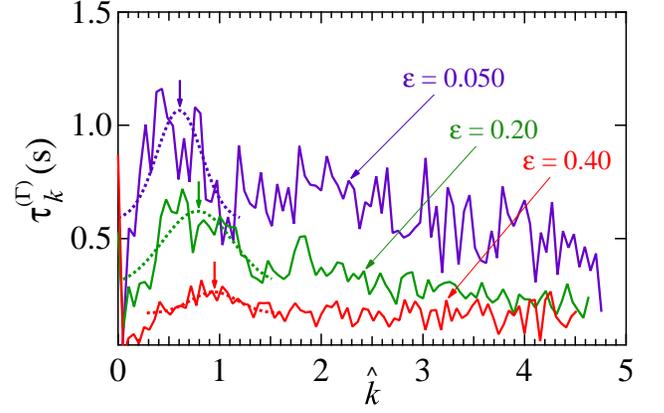}
\end{center}
	\caption{(color online)
		Wave-number dependences of the time scale $\tau^{(\Gamma)}_{k}$
		for several values of $\varepsilon$;
		$\varepsilon=0.050$ (purple), $0.20$ (green), and $0.40$ (red).
		Each dotted curve is obtained by a peak-fitting method and the vertical arrow points where $k_{\Gamma}$ is.
	}
\label{fig:tauG}
\end{figure}

\begin{figure}[!t]
\begin{center}
\includegraphics[width=0.99\linewidth]{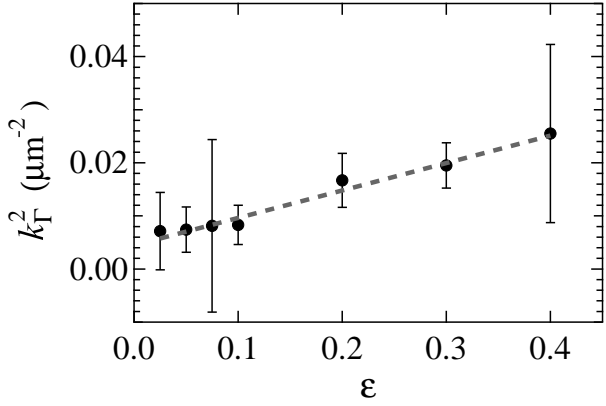}
\end{center}
\caption{%
	$\varepsilon$-dependence of $k_{\Gamma}$ at which $\tau^{(\Gamma)}_{k}$ has a local maximum.
	The gray dashed line marks the line fit, $k_{\Gamma}^2 = c_{0}+c_{1}\varepsilon$
	with $c_{0}=0.0045$ and $c_{1}=0.052$.
	The exponent of the power law is suggested from the SMT patch structure:
	 $\xi\sim\varepsilon^{-1/2}$.
}
\label{fig:kGsq}
\end{figure}

The characteristic time scale $\tau^{(\Gamma)}_{k}$, plotted as a function of normalized wave number in Fig.~\ref{fig:tauG},
has a weak peak in $\hat{k}<2$ for each $\varepsilon$.
Let $k_{\Gamma}$ denote the wave number at which $\tau^{(\Gamma)}_{k}$ has a peak,
where $k_{\Gamma}$ was determined by a method for multi-peak fitting so that roughness of curves was smoothened.
In expectation that the coherency in patch domains leads to the non-Markov memory effect,
the time scale $\tau^{(\Gamma)}_{k}$ should include features relating to the patch domains.
The length scale $\lambda_{\Gamma} = 2 \pi / k_{\Gamma}$ is several times larger than the diameter of an electroconvective roll.
In addition, the power law $\lambda_{\Gamma}\propto \varepsilon^{-1/2}$ is quantitatively reasonable as indicated in Fig.~\ref{fig:kGsq}.
Therefore, it follows that the dual relaxation caused by TFs originates from the patch structure.
The time scale rapidly decreases at larger length scales $k<k_{\Gamma}$,
indicating inter-patch dynamics does not affect the memory effect due to TFs.
Some work does support the relationship between the patch structure and the dual structure \cite{Tamura2002, Hidaka2010, Suzuki_tobe},
where the dynamics of a tagged particle in the SMT disorder can be divided into two types of mode:
one dominated by convective rolling within patch domains (i.e., intra-patch dynamics) 
and the other dominated by transfers with slow patch movements (i.e., inter-patch dynamics).

\begin{figure}[!t]
\begin{center}
\includegraphics[width=0.9\linewidth]{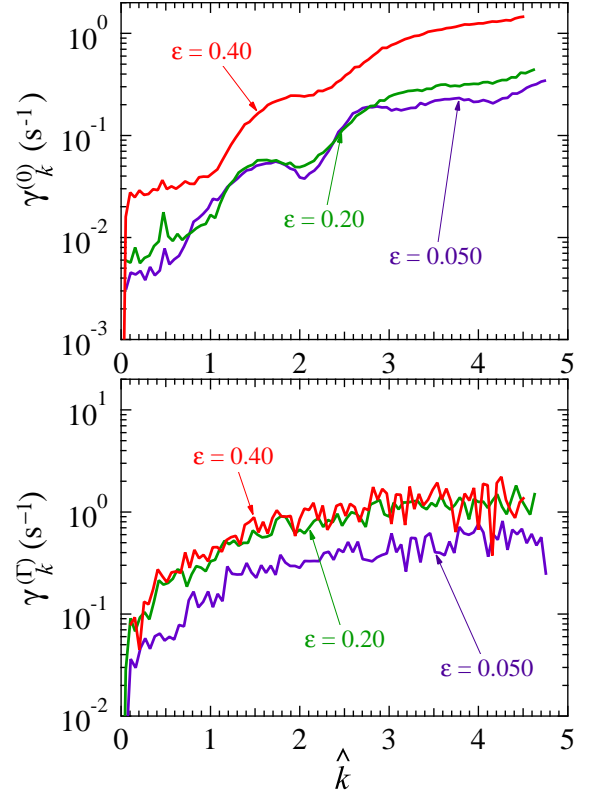}
\end{center}
	\caption{(color online)
		Dispersion relations for several $\varepsilon$ values;
		$\varepsilon=0.050$ (purple), $0.20$ (green), and $0.40$ (red).
		The top panel gives the bare friction coefficient $\gamma^{(0)}_{k}$,  
		and the bottom panel gives friction $\gamma^{(\Gamma)}_{k}$ caused by TFs.
	}
\label{fig:gamma}
\end{figure}

Figure \ref{fig:gamma} shows corresponding dispersion relations. 
The friction $\gamma^{(\Gamma)}_{k}$ is one order larger than the bare friction $\gamma^{(0)}_{k}$ near $k_{\Gamma}$,
but converges at large wave numbers.
A characteristic feature is a dip appearing near $\hat{k}\simeq 1$ in $\gamma^{(0)}_{k}$ but not in $\gamma^{(\Gamma)}_{k}$.
The bare friction $\gamma^{(0)}_{k}$ and the static friction $\gamma^{(\Gamma)}_{k}$ caused by TFs are explicitly related 
to the average rate in entropy production $\dot{S}_{k}$ \cite{Mori2001}.
Over macroscopic time scale, it is analytically represented as
\begin{equation}
	\dot{S}_{k} = k_{\text{B}}\left(\gamma^{(0)}_{k} + \gamma^{(\Gamma)}_{k}\right) = k_{\text{B}}\left/ \tau^{(\text{U})}_{k}\right.
\end{equation}
where $k_{\text{B}}$ is the Boltzmann constant.
The dip might reflect a law of minimum entropy production rate in electroconvection.

\begin{figure}[!t]
\begin{center}
\includegraphics[width=0.9\linewidth]{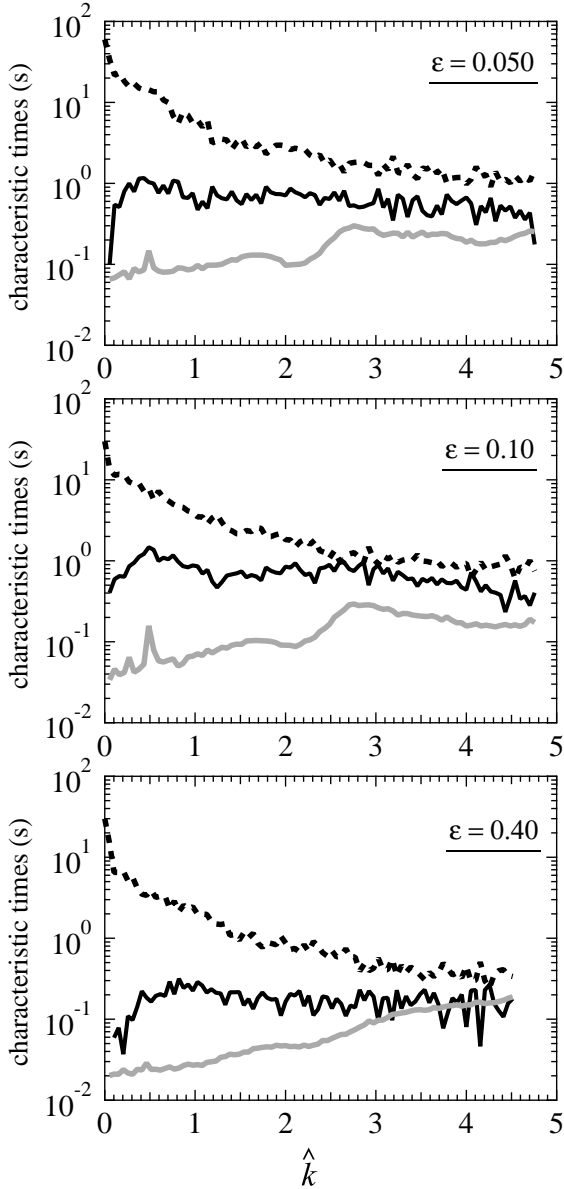}
\end{center}
	\caption{
		Plot of the characteristic times $\tau^{(\text{U})}_{k}$ (dashed line), 
		$\tau^{(\Gamma)}_{k}$ (dark solid line), 
		and $\tilde{\gamma}_{k}\tau^{(\Gamma)}_{k}$ (light solid line)
		for $\varepsilon = 0.050$ (top), $0.10$ (middle), and $0.40$ (bottom), respectively.
	}
\label{fig:comp_tau}
\end{figure}

The time scales, $\tau^{(\text{U})}_{k}$, $\tau^{(\Gamma)}_{k}$, and $\tilde{\gamma}_{k}\tau^{(\Gamma)}_{k}$, are compared 
in Fig.~\ref{fig:comp_tau} for several $\varepsilon$ values.
The correlation time $\tau^{(\text{U})}_{k}$ is much longer than $\tau^{(\Gamma)}_{k}$ at large length scales,
where the relaxation dynamics is approximately represented by a simple exponential.
The ratio $\tau^{(\text{U})}_{k}/\tau^{(\Gamma)}_{k}$ approaches unity with increasing $\hat{k}$,
signifying that the memory effect persists for relatively long times at short length scales.
Meanwhile, $\tilde{\gamma}_{k}\tau^{(\Gamma)}_{k}$ approaches $\tau^{(\Gamma)}_{k}$ at large $\hat{k}$,
implying that the duration of the early stage is shorter at small length scales.
Therefore, dual relaxation clearly appears in SMT at intermediate length scales,
where $\tau^{(\Gamma)}_{k}$ is not too short compared with $\tau^{(\text{U})}_{k}$ 
and the friction coefficient due to TFs is much larger than the bare friction, i.e., $\tilde{\gamma}_{k}\ll 1$.

Mori and Okamura have proposed the affinity hypothesis between the correlation and memory function \cite{Mori2009},
in which the correlation and memory functions have the same form of decay 
with different magnitudes and characteristic time scales.
The main prediction from the hypothesis is that, 
when $\tau^{(\text{U})}_{k}$ is almost the same as $\tau^{(\Gamma)}_{k}$,
decay of the correlation features a long-time tail with exponent $-3/2$ over all spatial dimensions.
Those time scales for SMT are almost the same for $\hat{k} \agt 3.0$,
as represented in Fig.~\ref{fig:comp_tau}.
Our modal correlation functions in the long-time region however have a large margin of error;
we cannot evaluate whether such long-time tails exist or not.

\section{Concluding remarks} \label{sec:summary}
We experimentally observed the modal autocorrelation function $\hat{U}_{k}(t)$ of pattern dynamics in SMT.
The modal relaxation dynamics featured a dual relaxation (Fig.~\ref{fig:dualstructure});
the correlation is well described by the time-reversible function at the early stage
and by a simple exponential relaxation at the late stage.
The corresponding time scales $\tau^{(\text{a})}_{k}$ and $\tau^{(\text{e})}_{k}$ obey dynamic scaling laws for the intra-patch scales (Fig.~\ref{fig:tauAandE}).

With an expectation that the patch structure generates memory effects due to the non-thermal fluctuations,
we employed the projection-operator formalism 
in which the memory function naturally arises as the autocorrelation function of the non-thermal fluctuations.
Indeed, we have verified the expectation;
the peak wave number $k_{\Gamma}$  (Fig.~\ref{fig:kGsq}) of the time scale $\tau^{(\Gamma)}_{k}$ (Fig.~\ref{fig:tauG}) suggests 
that the dual relaxation is caused from intra-patch dynamics.

Solving the evolution equation [Eq.~(\ref{EOM_modalcorr_prime})] derived in the projection-operator formalism,
we revealed two well-separated contributions for the non-thermal fluctuations in Fig.~\ref{fig:memboth_0100}:
rapidly-varying Markovian fluctuations and non-Markov TFs.
The former relates to the bare friction $\gamma^{(0)}_{k}$ and the latter to the memory function $\Gamma_{k}(t)$ [Eq.~(\ref{memory_function_SMT})].
SMT dynamics were shown analytically to separate into three stages (Table \ref{table:stage}):
the bare-friction stage,  $\tau \ll \tilde{\gamma}_{k}\tau^{(\Gamma)}_{k}$, where the relaxation is linear decay;
the early stage,  $\tilde{\gamma}_{k}\tau^{(\Gamma)}_{k} \ll \tau \ll \tau^{(\Gamma)}_{k}$, 
where the relaxation is time-reversible algebraic decay [Eq.~\eqref{algebraic_decay}];
and the late stage, $\tau^{(\Gamma)}_{k} \ll \tau$, where the relaxation is simple exponential decay [Eq.~\eqref{exponential_decay}].
Further, by the variables defined in the theory (i.e., $\gamma^{(0)}_{k}$, $\gamma^{(\Gamma)}_{k}$, and $\tau^{(\Gamma)}_{k}$), 
we have analytically specified time scales obtained in experiments:
$\tau^{(\text{a})}_{k}$ [Eq.~\eqref{theory_tau_a}], $\tau^{(\text{e})}_{k}$ [Eq.~\eqref{theory_tau_e}], and $\tau^{(\text{U})}_{k}$ [Eq.~\eqref{relation_tau}].

Comparison between the characteristic time scales (Fig.~\ref{fig:comp_tau})
clarified the nature of the dual structure in SMT.
It clearly appears at intermediate length scales
where the characteristic time $\tau^{(\text{U})}_{k}$ of the modal relaxation is not too short 
compared with that $\tau^{(\Gamma)}_{k}$ of the memory function due to TFs.
Also, the friction coefficient $\gamma^{(\Gamma)}_{k}$ due to TFs is much larger than the bare friction $\gamma^{(0)}_{k}$.

The physical origin of the Markovian non-thermal fluctuations is still an open question.
The dip near $\hat{k}\simeq 1$ in Fig.~\ref{fig:gamma} suggests 
that the Markov contribution contains spatial features relating to electroconvection.
That is, the bare friction consists of not only thermal fluctuations but also rapid variation in non-thermal ones.
Note that the Markovian non-thermal fluctuations seem not to affect the SMT pattern dynamics,
but acts as a trigger for the onset of SMT.
If non-thermal Markov fluctuations played a role in pattern dynamics,
the modal correlation would break time-reversal symmetry even at the early stage;
nevertheless, early-stage relaxation is invariant under time reversal [Eq.~\eqref{algebraic_decay}].

Besides SMT,
several types of spatiotemporal chaos appear in the electroconvection of the liquid crystal systems.
For example, instead of the homeotropic anchoring, one can apply the memory-function procedure to the planner anchoring as the boundary condition
in which the rotational symmetry is not alive in the $x$--$y$ plane.
Our approach proposed in this paper can pave the way to understand them universally from a viewpoint of the non-thermal fluctuations relating to the pattern.
This  will be discussed elsewhere.

\begin{acknowledgments}
The authors gratefully acknowledge Dr.~Makoto Okamura (Kyushu University) and Dr.~Kazumasa A.~Takeuchi (the University of Tokyo) for productive discussions.
This work was partially supported by KAKEHNI (Nos.~21340110 and 24540408),
a Grant-in-Aid for Scientific Research on Innovative Areas---"Emergence in Chemistry"~(No.~20111003),
and the JSPS Core-to-Core Program "International research network for non-equilibrium dynamics of soft matter."
\end{acknowledgments}

\appendix
\section{Projection-Operator Formalism} \label{appendix:NPOF}
Physically, a macroscopic system can be divided into slowly varying behavior described by a set of macroscopic variables $\{A_{i}\}$
and rapidly-varying terms \cite{Green1952, Kubo1966}.
The projection-operator formalism mathematically allows us to separate these two dynamics.

Using the linear projection operator,
Mori mathematically derived a generalized linear Langevin equation containing a memory function \cite{Mori1965}.
The memory function is represented as temporal correlation of the fluctuations $r_{i}(t)$,
where $\left<r_{i}(t+\tau)A_{j}(t)\right>=0$.
The generalized linear Langevin equation can be employed for not only equilibrium but also non-equilibrium systems.
However, in the \textit{linear} projection scheme, the fluctuation term $r_{i}(t)$ is orthogonal only to the linear functions of $\{A_{i}\}$
and can consist of not only the microscopic degrees of freedom but also fluctuations from nonlinear terms of $\{A_{i}\}$.
Such \textit{non-thermal} fluctuations are possibly relevant in nonequilibrium systems \cite{Kawasaki1970, Zwanzig1972}.

Zwanzig formulated the generalized nonlinear Langevin equation 
by the \textit{nonlinear} projection operator \cite{Zwanzig1960, Zwanzig1961} as
\begin{equation}
	\dot{A}_{i}(t) = v_{i}(t) + J_{i}(t) + R^{(0)}_{i}(t),
	\label{nonlinear_GLE}
\end{equation}
where $v_{i}(t)$ denotes the streaming term including nonlinear reversible terms and $J_{i}(t)$ an irreversible term.
The fluctuation term $R^{(0)}_{i}(t)$ satisfies $$\left<R^{(0)}_{i}(t+\tau)\mathcal{F}(A(t))\right>=0$$ 
for an arbitrary function $\mathcal{F}(A)$ of a set of macroscopic variables, $\{A_{i}\}$.
The nonlinearity of $\{A_{i}\}$ is included in $v_{i}(t)$ and $J_{i}(t)$,
and one can extract the linear part of $\{A_{i}\}$ from $v_{i}(t)$ and $J_{i}(t)$ using the \textit{linear} projection-operator method.
Mori and Fujisaka derived the evolution equation for the correlation function $U_{ij}(\tau) \coloneqq \left<A_{i}(t+\tau)A^{*}_{j}(t)\right>$ as \cite{Mori1973, Mori2001}
\begin{equation}
	\frac{\partial U_{ij}(\tau)}{\partial \tau}=\sum_{l}\text{i}\omega_{il}U_{lj}(\tau)-\sum_{l}\int_{0}^{\tau}{\rm d}\tau^{\prime} \Gamma^{\prime}_{il}(\tau-\tau^{\prime})U_{lj}(\tau^{\prime}),
	\label{nonlinear_evolution_corr}
\end{equation}
where $\omega_{ij}$ denotes the mechanical coefficient and $\Gamma^{\prime}_{il}(\tau)$ the memory function.
Note that Eq.~(\ref{nonlinear_evolution_corr}), 
which relates the memory $\Gamma^{\prime}(\tau)$ to the non-thermal fluctuations,
is of the same form as the evolution equation derived in the linear projection-operator formalism;
therefore, from a physical point of view, the obtained memory functions should be checked to see 
whether the memory effect originates from non-thermal fluctuations or not.
If the characteristic time scale for the correlation of $R^{(0)}_{i}(t)$ is extremely short,
then $R^{(0)}_{i}(t)$ satisfies; 
\begin{equation}
	\left<R^{(0)}_{i}(t+\tau)R^{(0)}_{j}(t)\right> \simeq \gamma^{(0)}_{ij}\delta(\tau),
	\label{bare_friction}
\end{equation}
where $\gamma^{(0)}_{ij}$ denotes the bare friction coefficient.
In addition, if the fluctuations extracted from the nonlinear terms do not correlate with $R^{(0)}_{i}(t)$,
the memory function due to the non-thermal fluctuations can be divided into two different terms,
\begin{equation}
	\Gamma^{\prime}_{ij}(\tau) = 2\gamma^{(0)}_{ij}\delta(\tau)+\Gamma_{ij}(\tau),
	\label{memory_function}
\end{equation}
where $\Gamma_{ij}(\tau)$ is regarded as the temporal correlation of the non-Markov contribution of the non-thermal fluctuations.

%

\end{document}